  \documentclass[conference]{IEEEtran}

   \ifCLASSINFOpdf   
\else
\fi
\usepackage{times,epsfig,latexsym,amssymb,psfrag}
\usepackage{paralist,citesort}
\usepackage{booktabs}
\usepackage{subcaption}
\usepackage{eurosym}

\usepackage{mathrsfs}
\usepackage{enumitem}
\usepackage[ruled,vlined]{algorithm2e}
\usepackage{graphicx}
\usepackage{epstopdf} %converting to PDF
\usepackage{amsmath}
\usepackage{mathtools}

 \newcommand{\qed}{\nobreak \ifvmode \relax \else
 	\ifdim\lastskip<1.5em \hskip-\lastskip
 	\hskip1.5em plus0em minus0.5em \fi \nobreak
 	\vrule height0.75em width0.5em depth0.25em\fi}

\begin{document}
 \title{Bitcoin Price Prediction: An ARIMA Approach}

\author{Amin Azari\\
KTH Royal Institute of Technology, Email: aazari@kth.se} 
 \maketitle

\begin{abstract}
 Bitcoin is considered  the most valuable currency in the world. Besides being highly valuable, its value has also experienced a steep increase, from around 1 dollar in 2010 to around 18000 in 2017. Then, in recent years, it  has attracted considerable attention in a diverse set of  fields, including  economics  and computer science. The former mainly focuses on studying how it affects the market, determining reasons behinds its price fluctuations, and predicting its future prices.  The latter  mainly focuses on its vulnerabilities, scalability, and other techno-crypto-economic issues. Here, we aim  at revealing the usefulness of traditional autoregressive integrative moving average (ARIMA) model in predicting the future value of bitcoin by analyzing the price time series in a 3-years-long time period.  On the one hand, our empirical studies  reveal that  this simple scheme is efficient in sub-periods in which the behavior of the time-series is almost unchanged, especially when it is used for short-term prediction, e.g. 1-day. On the other hand, when  we try to train the ARIMA model to a 3-years-long period, during which the bitcoin price has experienced different behaviors, or when we try to use it for a long-term prediction, we observe that it introduces large prediction errors.  Especially, the ARIMA model  is unable to capture the sharp fluctuations in the price, e.g. the volatility at the end of 2017. Then, it calls for more features to be extracted and used along with the price for a more accurate prediction of the price. We have further investigated the bitcoin price prediction using an ARIMA model, trained over a large dataset, and a limited test window of the bitcoin price, with length $w$, as inputs.  Our study sheds lights on the interaction of the prediction accuracy, choice of ($p,q,d$), and window size $w$.
\end{abstract}
 
\IEEEpeerreviewmaketitle
 
 %%------------------------------------------------------------
 
 \section{Introduction}
Traditionally the richest people in a society have an average age of more than 60 years \cite{age}. The exponential rise in the information and communications technology (ICT) have not only changed many of our beliefs, habits and traditions, but  also it has introduced a new wave of young billionaires. By introduction of bitcoin in 2009 with an initial value of around one dollar, no one predicts that in 8 years it will pass all previous records, and will reach to the unbelievable value of 18000 dollars. Due to such a rapid increase in the price of bitcoin and its following price correction, i.e. fall down of the price after the big jump, many economists and  computer scientists tried to explore the  time series of the bitcoin price \cite{amjad,bit,emp}.  Our focus in this letter is to investigate application of the traditional autoregressive integrative moving average (ARIMA) model in prediction of bitcoin price. ARIMA has been used extensively in prediction of stationary datasets in the literature, e.g. refer to \cite{marj} and its references. Application of ARIMA in prediction of bitcoin price has been recently investigated in \cite{oona}, and it has been shown that a Recurrent Neural Network model (RNN), augmented by additional data like Tweeter's hashtags can significantly outperform the performance, especially in long-term predictions. The interesting discovery of \cite{oona} consists in the fact that in short time periods, ARIMA   outperforms RNN, which both of them work on a single input, i.e. the price of bitcoin. Regarding the computational complexity associated with the neural networks, and the explainability of results achieved by using ARIMA models,  here we aim at further investigation of prediction of bitcoin price using ARIMA model for 1-day prediction of the bitcoin price. More specifically, we aim at  investigating the couplings among the length of training period, the choice of ARIMA parameters $(p,q,d)$, and the length of time window that the prediction is carried out over it, i.e. the bitcoin price for the day after the window is predicted. The remainder of this letter is as follows. In the next section, we briefly present the system model and formulate the problem. In Section III, we  present our approach for solving the problem. Section IV presents the performance evaluation results. Concluding remarks are given in Section V.

 \section{System Model}
Fig. \ref{raw} represents the closing price of bitcoin (in dollars) for a period of 3 years, starting from the 2015-09-01. The research problem consists in training an ARIMA ($p,q,d$) predictor over this data set, such that given a time window of length $w$, the prediction error for 1 day after the window is minimized. Here,  $p$ $q$, $d$, and coefficients of the ARIMA predictor are the degrees of freedom.

 \section{ARIMA-based Price Prediction}
This section includes our approach for solving this problem. 
 \subsection{ Preprocessing of the Dataset: Making data stationary}
\begin{figure}[t!]
        \centering
                \includegraphics[trim={2.5cm 0 3.5cm  1cm},clip,width=3.5in]{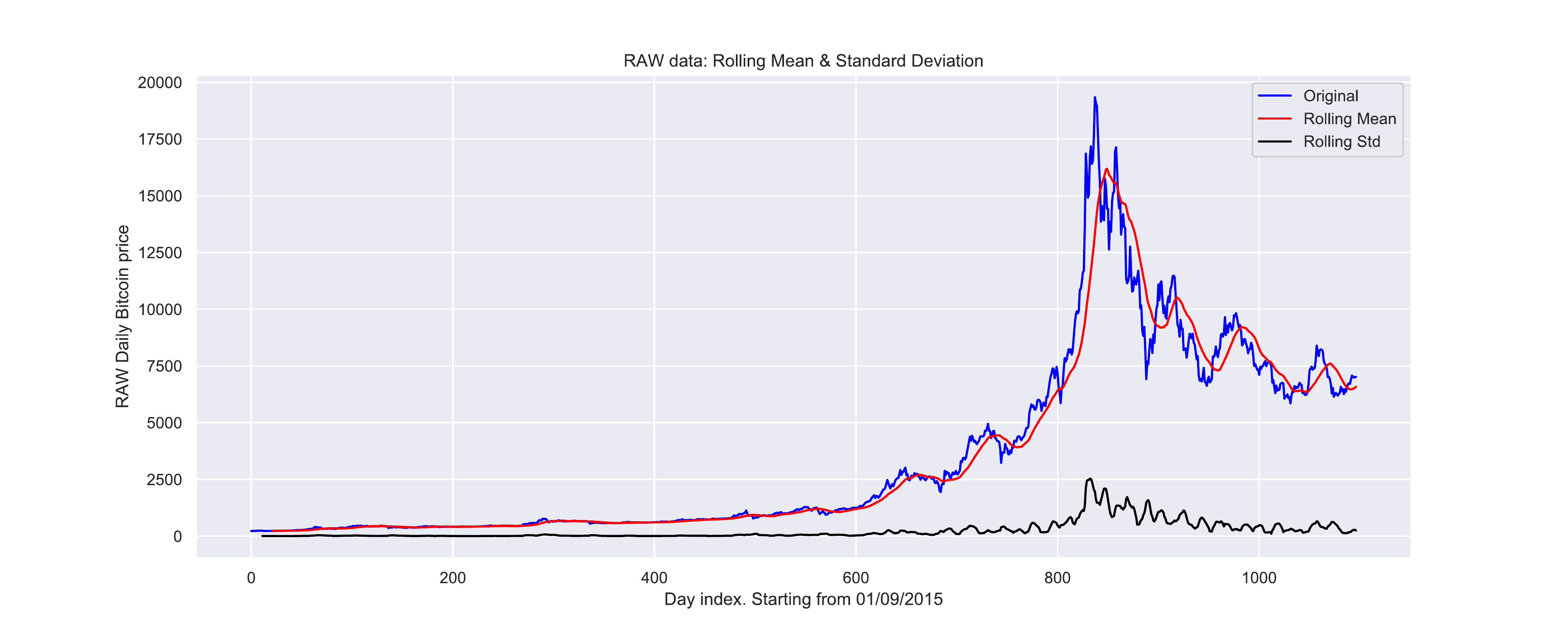}
     \caption{Analysis of the raw data}
                     \label{raw}
 \end{figure}

\begin{figure}[t!]
        \centering
                \includegraphics[trim={3cm 0 3.5cm  1cm},clip,width=3.5in]{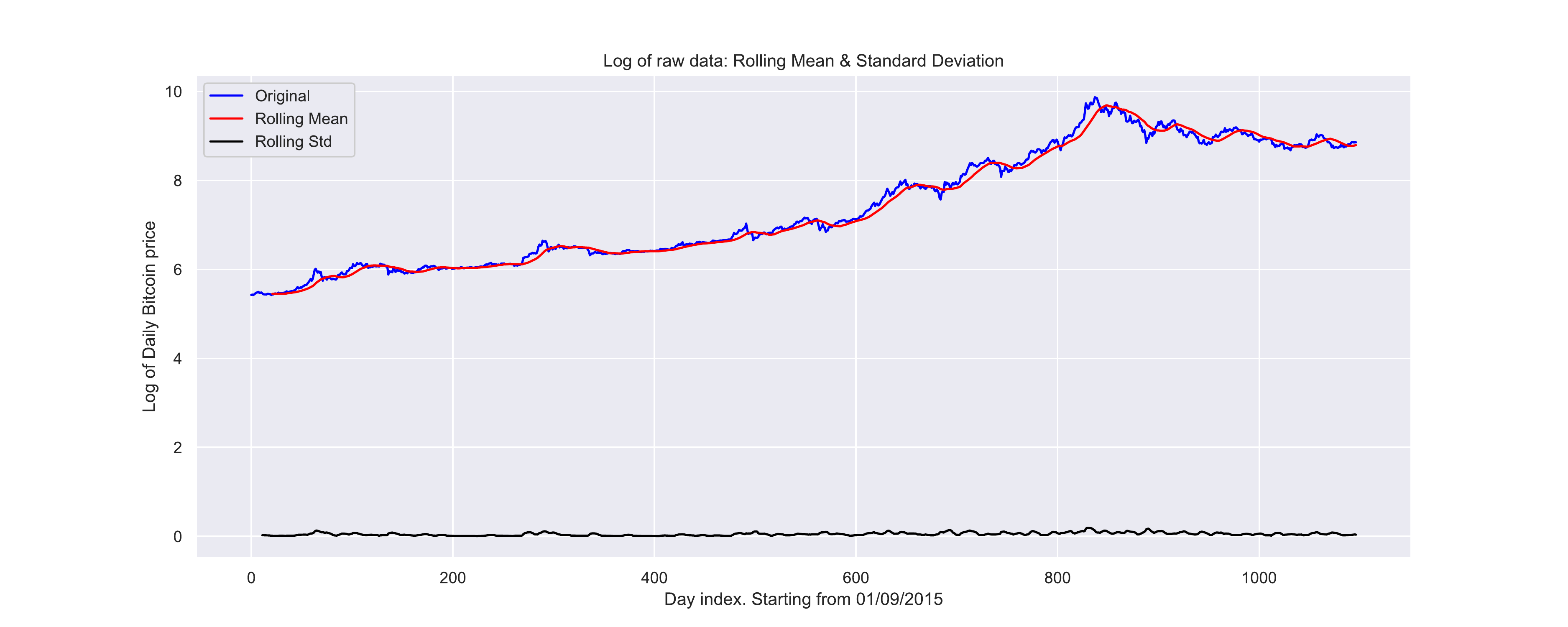}
     \caption{Analysis of the log of raw data}
                     \label{log}
 \end{figure}

\begin{figure}[t!]
        \centering
                \includegraphics[trim={3cm 0 3.5cm  1cm},clip,width=3.5in]{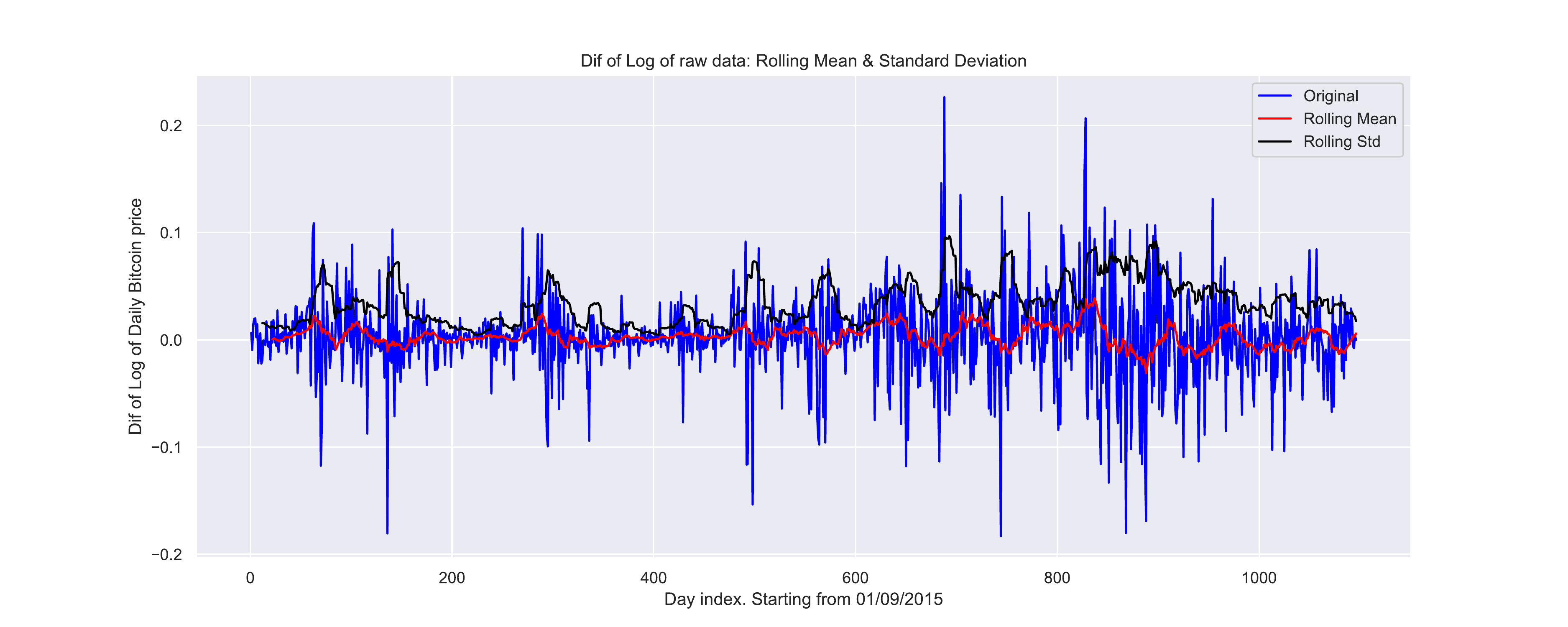}
     \caption{Analysis of the differential of log of the raw data}
                     \label{dif}
 \end{figure}

\begin{figure}[t!]
        \centering
                \includegraphics[trim={3cm 0 3.5cm  1cm},clip,width=3.5in]{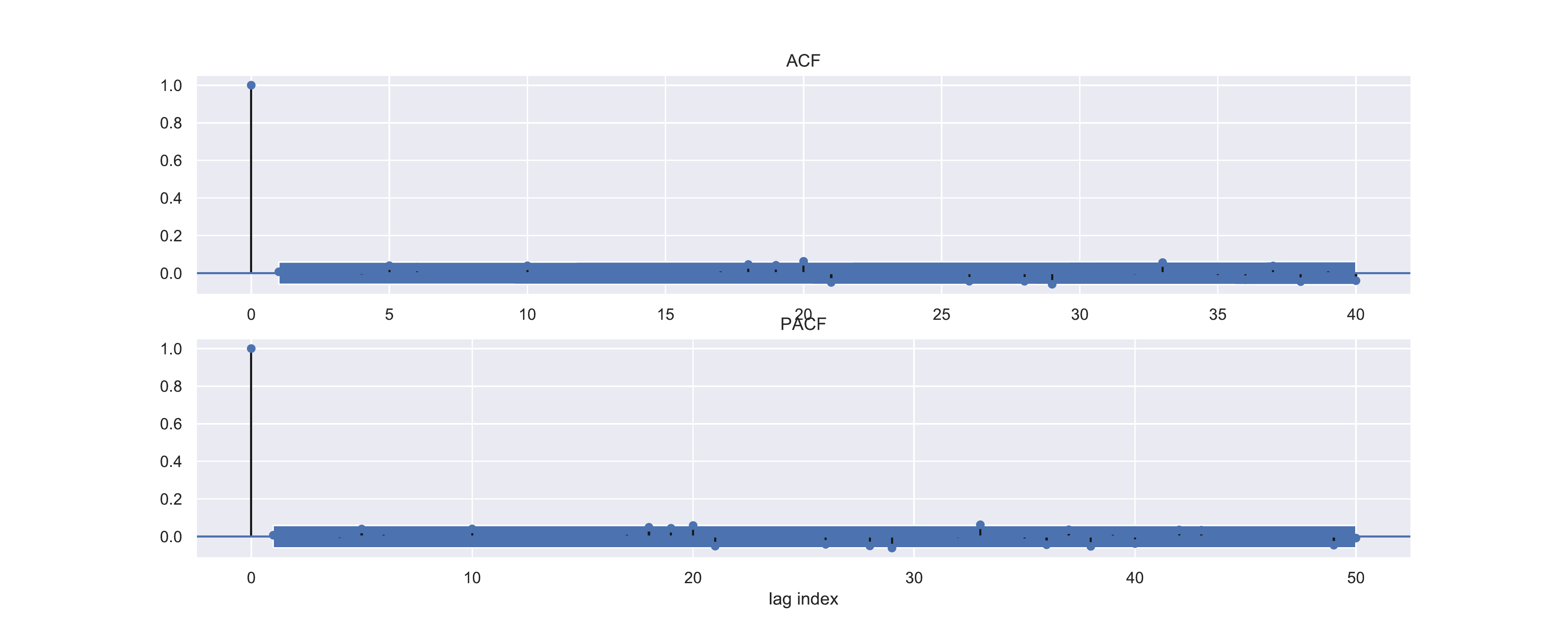}
     \caption{Analysis of ACT and PACF curves}
                     \label{pacf}
 \end{figure}

First of all, using ARIMA models requires dataset to be stationary. A first look at the dataset (Fig.\ref{raw}) shows that it has an exponential trend. Taking the logarithm of the data (Fig. \ref{log}), we still observe an increasing trend in the data. Especially, the Autocorrelation and partial autocorrelation functions (ACT/PACF), which have been omitted here for the brevity of test represents a high coupling between prices, e.g in order of 80 days. This motivates us to benefit from the differential transform. Taking the first order differential transform (Fig. \ref{dif}), one observes that the trend in the data has been removed. The stationarity test also shows that the $p$-value is very close to zero, and is less than the  critical values. Now, we are sure that the dataset has become stationery. While traditionally we benefit from the ACF/PACF for estimating the $p$ and $q$ values, the ACF/PACF figures (Fig. \ref{pacf}) show that the there is a weak ARMA relationship in the transformed data. But, why does this happen? This is mainly due to the fact that the bitcoin price has different behaviors in the first and second half of the time span, and a single ARIMA model cannot well fit to the whole time period. But here we don't break the dataset and continue with finding the best ARIMA model which fits the whole time span, however, we expect that its performance will be weak\footnote{The preprocessing of data for making it stationary has been implemented in Python, and is available in our repository supporting this document (https://github.com/AminAzari/ARIMA), with the name of part1.py}. We implement two approaches for identifying ($p,g,d$), including (i) the one which minimizes the residual sum of squares (RSS) in the model fitting phase, and (ii) the one that minimizes the mean square error (MSE) in the prediction phase. These two approaches have been implemented in Python, and are available in our repository supporting this document. The related files to RSS and MSE approaches are part2.py and part3.py respectively.  The former aims at performing a grid search over feasible values of $p$, $q$, and $d$, deriving the RSS from fitting dataset to the ARIMA model, and reporting the one which minimizes the RSS. The latter aims at performing a grid search for finding the model which can generalize and minimize the prediction error. In other words, here we train the ARIMA model based on the dataset. Then we generate random time windows of length $w$, and test each ARIMA model for predicting the day after the window.   Then, we select the model with the lowest prediction error.

In the next section, we compare these two schemes and present the suggested ARIMA models by each scheme.


\section{Performance Evaluation}
\begin{figure}[t!]
        \centering
                \includegraphics[trim={3cm 0 3.5cm  1cm},clip,width=3.5in]{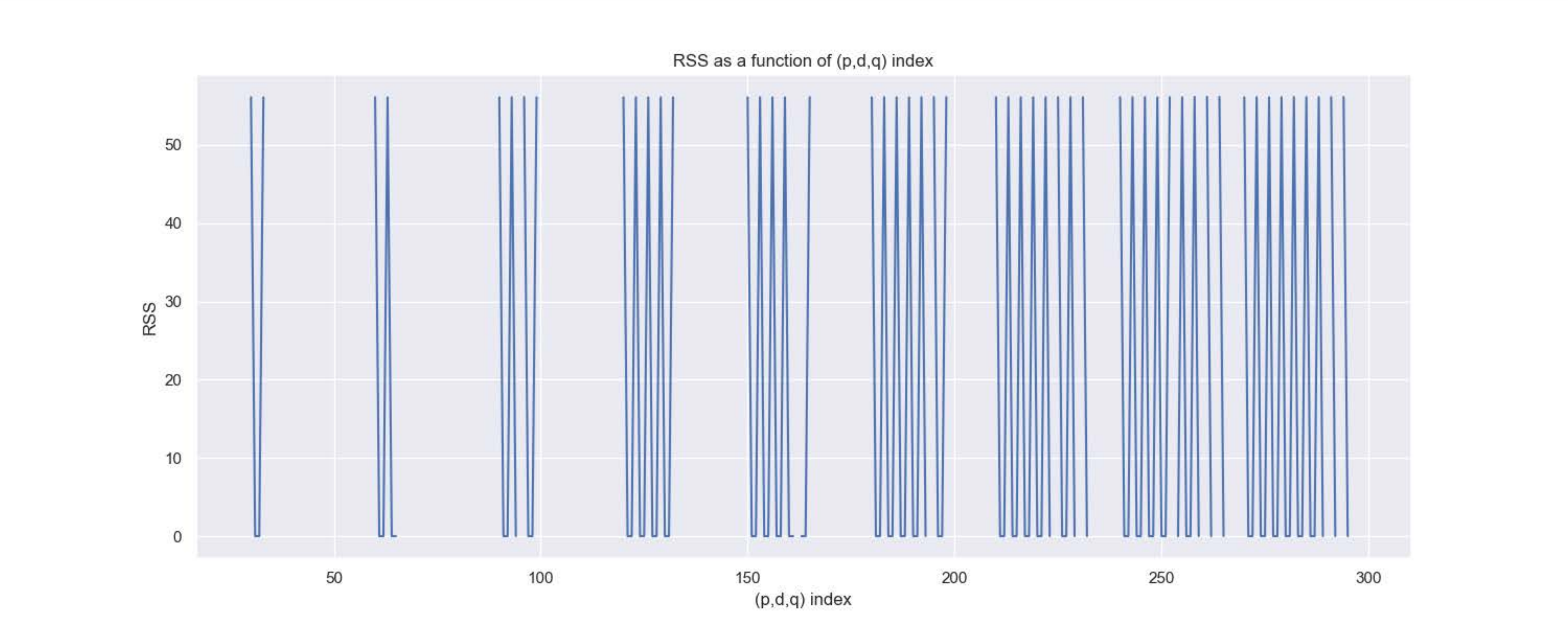}
     \caption{RSS as a function of ($p,q,d$) index for $p\ge q$. The minimum reported RRS is 0.002, which is achieved by (8,8,1). }
                     \label{rss}
 \end{figure}

\begin{figure}[t!]
        \centering
                \includegraphics[trim={3cm 0 3.5cm  1cm},clip,width=3.5in]{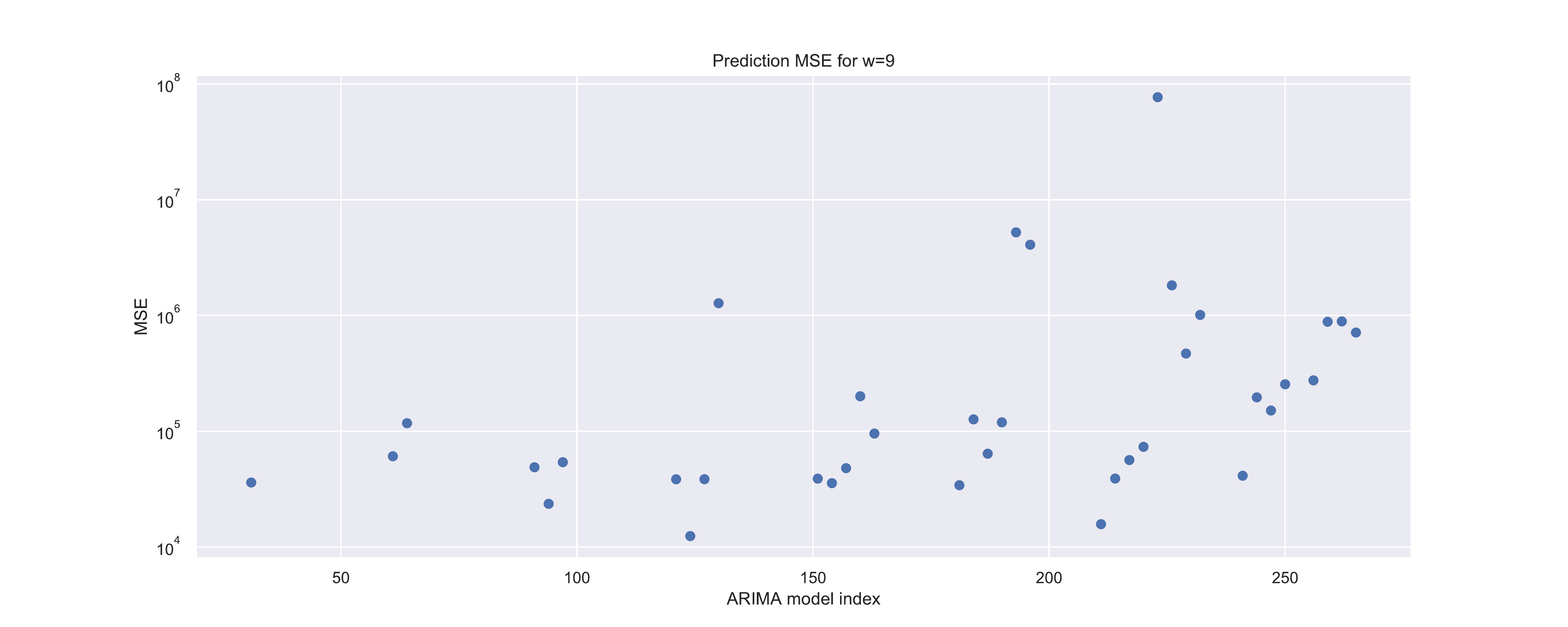}
     \caption{MSE results for $w$=9 vs. different ARIMA schemes}
                     \label{mse}
 \end{figure}

\begin{figure}[t!]
        \centering
                \includegraphics[trim={3cm 0 3.5cm  1cm},clip,width=3.5in]{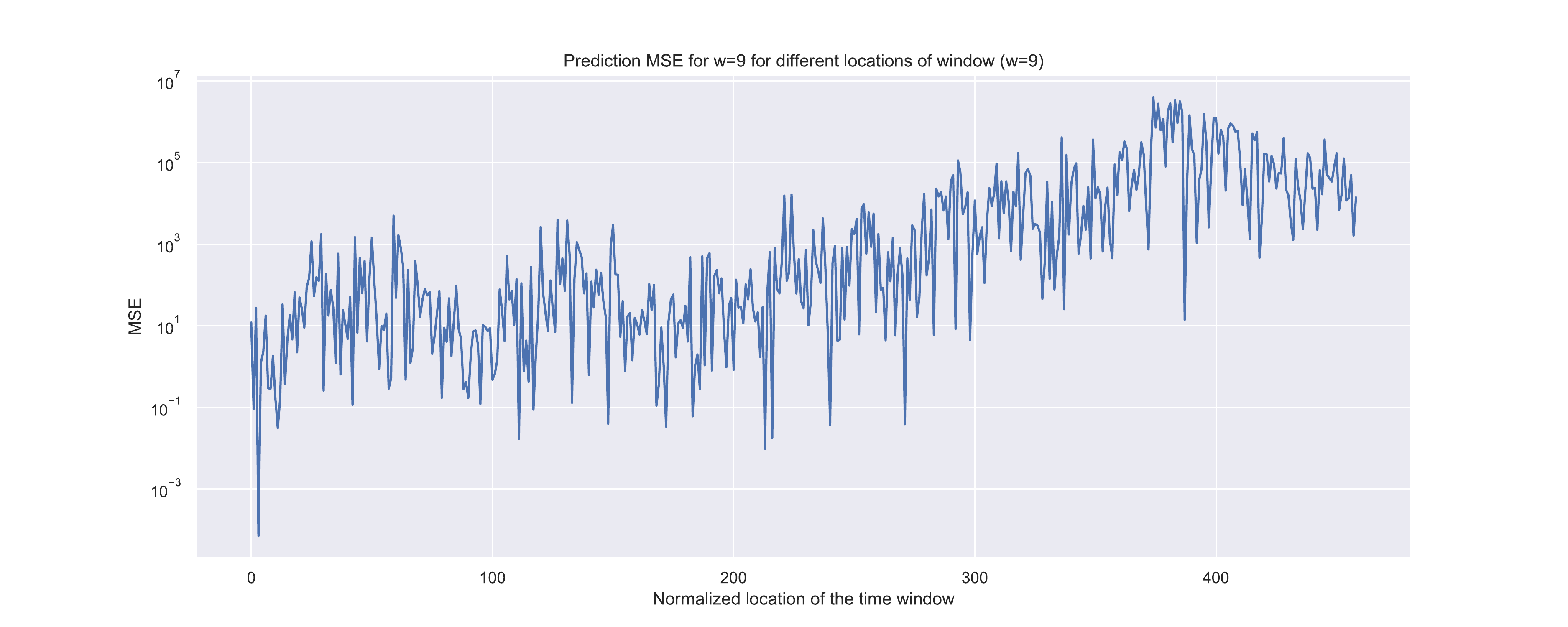}
     \caption{MSE results for $w$=9 vs. different locations of the time window for ARIMA(4,1,1)}
                     \label{msel}
 \end{figure}

In this section we  compare the performance of the RSS-based and the MSE-based approaches. Fig. \ref{rss} represents the RSS results as a function of the ARIMA ($p,q,d$) model index. The index of each ($p,q,d$) tuple is determined from a $for$ loop on $p\in\{0,\cdots,9\}$, $q\in\{0,\cdots,9\}$, and $d\in\{0,\cdots,2\}$.  E.g.,   (0,0,0) is shown by 0 and  (0,0,1) is shown by 1. The NaN values correspond to the cases where $p\le q$, or when ($p,q,d$) cannot make an ARIMA predictor. The minimum reported RRS is 0.002, which is achieved by (8,8,1), i.e.  index of 265. Fig. \ref{mse} represents  the MSE results for different ($p,q,d$) configurations, when the test window size is $w=9$. The location of time window has been selected randomly 50 times, and for each selections, 40 representations have been carried out. One sees here that the ARIMA(8,8,1), which achieved the minimum RSS values, achieves an MSE, which is almost 100 times more than the minimum achievable MSE (through ARIMA(4,2,1) with index 125 as shown in Fig. \ref{mse}). To get further insight  into the source of the large reported MSEs in Fig. \ref{mse}, we also depict the level of MSE for different locations of the time window, i.e. for predictions in different dates, ranging from September 2015 to 2018. Fig. \ref{msel} represents the MSE values as a function of day number, starting from 01/09/2015.  One sees that the maximum experienced MSE is related to the volatility in the bitcoin price in 2017. On the other hand, from 2015-early 2017, in which bitcoin price follows a more predictable trend, the predictions error is much less than the prediction error in 2017-2018 period.

\begin{table}[]
\caption{Optimized ARIMA models as a function of  $w$. The time window location is selected in  the whole time span.}\label{t1}
\begin{tabular}{llllll}
$w$ & \begin{tabular}[c]{@{}l@{}}RSS \\ mod. ind.\end{tabular} & \begin{tabular}[c]{@{}l@{}}RSS\\ model\end{tabular} & \begin{tabular}[c]{@{}l@{}}MSE\\ mod. index\end{tabular} & \begin{tabular}[c]{@{}l@{}}MSE\\ model\end{tabular} & Avg. MSE \\
\hline
2 & 31  & (1,1,0)            & 31  & (1,1,0)            & 118 K    \\
3 & 64  & (2,1,1)            & 31  & (2,1,0)            & 65K      \\
5 & 130 & (4,1,3)            & 61  & (2,1,0)            & 53K      \\
6 & 130 & (4,1,3)            & 94  & (3,1,1)            & 28K      \\
9 & 265 & (8,1,8)            & 124 & (4,1,1)            & 16K     
\end{tabular}

\end{table}

\begin{table}[]
\caption{Optimized ARIMA models as a function of $w$.  The time window location is selected in first half of the time span.}\label{t2}

\begin{tabular}{llllll}
$w$ & \begin{tabular}[c]{@{}l@{}}RSS \\ mod. ind.\end{tabular} & \begin{tabular}[c]{@{}l@{}}RSS\\ model\end{tabular} & \begin{tabular}[c]{@{}l@{}}MSE\\ mod. index\end{tabular} & \begin{tabular}[c]{@{}l@{}}MSE\\ model\end{tabular} & Avg. MSE \\
\hline
2 & 31  & (1,1,0)            & 31  & (1,1,0)            & 167    \\
3 & 64  & (2,1,1)            & 31  & (1,1,0)            & 45      \\  
\end{tabular}
\end{table}

After investigating the impact of locations of time window in the prediction error,  we investigate the impact of length of time window on the prediction error. Table \ref{t1} represents several values of the window size, the suggested ARIMA model by RSS, the MSE minimizing ARIMA model, and the respective minimum achievable MSE value by the MSE-minimizing ARIMA model.   From this table, one can observe that the increase in the $q$ by increasing $w$ is small. This is due to the fact that using moving average for prediction requires initialization of the model with random prediction error, and the impact of this initializations disappear slowly by increasing the window size.  On the other hand, one sees that the MSE-minimizing ARIMA model accepts some level of RSS in fitting to the data in order to be able to capture fluctuations in the price, i.e. the RSS-minimizing model and MSE-minimizing models are not in match, and the RSS-minimizing model is over-fitted to the data. To finalize the discussion, we have further reported in Table \ref{t2} the performance evaluation results for a case in which, the ARIMA model is trained using the whole data set, but it is tested for time windows in the first half of the dataset. One sees that the level of prediction error is very low, in comparison with the prior case. 

 \section{Conclusion}
In this letter, we have investigated bitcoin price prediction by using an ARIMA model. Towards this end, at first we have preprocessed data to make it stationary, and then, have searched over feasible ($p,q,d$) tuples for finding the ARIMA model which minimizes the MSE of prediction. Our results indicate that the  bitcoin price prediction using its closing price history could results in large MSE values due to bitcoin's price vulnerability to high jumps and fall-downs. on the other hand,  the results confirm that the  ARIMA model could be still used for price prediction in sub-periods of the timespan, i.e. by dividing the timespan to several time-spans over which, dataset has a unique trend. Furthermore, we have investigated the impact of location of the test time window and its length on the achieved MSE in price prediction. Specifically, we show that by increase in the time window from 2 to 9, the average MSE of prediction error could be decreased from 118000 to 16000.

 \ifCLASSOPTIONcaptionsoff
  \newpage
\fi

\bibliographystyle{IEEEtran}
\bibliography{report}

\begin{thebibliography}{1}
\providecommand{\url}[1]{#1}
\csname url@samestyle\endcsname
\providecommand{\newblock}{\relax}
\providecommand{\bibinfo}[2]{#2}
\providecommand{\BIBentrySTDinterwordspacing}{\spaceskip=0pt\relax}
\providecommand{\BIBentryALTinterwordstretchfactor}{4}
\providecommand{\BIBentryALTinterwordspacing}{\spaceskip=\fontdimen2\font plus
\BIBentryALTinterwordstretchfactor\fontdimen3\font minus
  \fontdimen4\font\relax}
\providecommand{\BIBforeignlanguage}[2]{{%
\expandafter\ifx\csname l@#1\endcsname\relax
\typeout{** WARNING: IEEEtran.bst: No hyphenation pattern has been}%
\typeout{** loaded for the language `#1'. Using the pattern for}%
\typeout{** the default language instead.}%
\else
\language=\csname l@#1\endcsname
\fi
#2}}
\providecommand{\BIBdecl}{\relax}
\BIBdecl

\bibitem{age}
\BIBentryALTinterwordspacing
Wikipedia. (2018) The world's billionaires. [Online]. Available:
  \url{https://en.wikipedia.org}
\BIBentrySTDinterwordspacing

\bibitem{amjad}
M.~Amjad and D.~Shah, ``Trading bitcoin and online time series prediction,'' in
  \emph{NIPS 2016 Time Series Workshop}, 2017, pp. 1--15.

\bibitem{bit}
S.~McNally, J.~Roche, and S.~Caton, ``Predicting the price of bitcoin using
  machine learning,'' in \emph{IEEE 26th Euromicro International Conference on
  Parallel, Distributed and Network-based Processing (PDP)}, 2018, pp.
  339--343.

\bibitem{emp}
H.~Jang and J.~Lee, ``An empirical study on modeling and prediction of bitcoin
  prices with bayesian neural networks based on blockchain information,''
  \emph{IEEE Access}, vol.~6, pp. 5427--5437, 2018.

\bibitem{marj}
P.~J. Brockwell, R.~A. Davis, and M.~V. Calder, \emph{Introduction to time
  series and forecasting}.\hskip 1em plus 0.5em minus 0.4em\relax Springer,
  2002, vol.~2.

\bibitem{oona}
J.~Rebane, I.~Karlsson, S.~Denic, and P.~Papapetrou, ``{Seq2Seq} {RNNs} and
  {ARIMA} models for cryptocurrency prediction: A comparative study,'' in
  \emph{{ FinTech-KDD}}, 2018.

\end{thebibliography}

\end{document}